\documentclass[conference]{IEEEtran}
\IEEEoverridecommandlockouts

\usepackage{cite}
\usepackage{amsmath,amssymb,amsfonts}
\usepackage{algorithmic}
\usepackage{graphicx}
\usepackage{textcomp}
\usepackage{xcolor}
\usepackage{textgreek}
\usepackage{rotating}
\usepackage{adjustbox}
\usepackage{booktabs}
\usepackage{pdflscape}
\usepackage{hyperref}

\usepackage[locale=US, separate-uncertainty=true]{siunitx}

\DeclareSIUnit{\samples}{Sa}
\DeclareSIUnit{\dBm}{\text{dBm}}
\DeclareSIUnit{\mAh}{\text{mAh}}

\def\BibTeX{{\rm B\kern-.05em{\sc i\kern-.025em b}\kern-.08em
    T\kern-.1667em\lower.7ex\hbox{E}\kern-.125emX}}

\begin{document}

\title{Precise Low-Current Measurement Techniques for IoT Devices: A Case Study on MoleNet}

\author{\IEEEauthorblockN{Julian Block}
\IEEEauthorblockA{\textit{Sustainable Comm. Networks} \\
\textit{University of Bremen}\\
Bremen, Germany \\
ju\_bl@uni-bremen.de}
\and
\IEEEauthorblockN{Andreas Könsgen}
\IEEEauthorblockA{\textit{Sustainable Comm. Networks} \\
\textit{University of Bremen}\\
Bremen, Germany \\
ajk@comnets.uni-bremen.de}
\and
\IEEEauthorblockN{Jens Dede}
\IEEEauthorblockA{\textit{Sustainable Comm. Networks} \\
\textit{University of Bremen}\\
Bremen, Germany \\
jd@comnets.uni-bremen.de}
\and
\IEEEauthorblockN{Anna Förster}
\IEEEauthorblockA{\textit{Sustainable Comm. Networks} \\
\textit{University of Bremen}\\
Bremen, Germany \\
anna.foerster@uni-bremen.de}
} 

\maketitle

\begin{abstract}
Power consumption is a crucial aspect of IoT devices which often have to run on a battery for an extended period of time. Therefore, supply current measurements are crucial before deploying a device in the field. Multimeters and oscilloscopes are not well suited when it comes to measuring very small currents which occur e.g. when an IoT device is in sleep mode. In this report, we compare dedicated source measurement units (SMUs) which allow to measure very small currents with high precision. As an application example, we demonstrate current measurements on our MoleNet IoT sensor board.
\end{abstract}


\section{Introduction}
The Internet of Things is rapidly gaining popularity, IoT devices are meanwhile deployed in various types of applications. In many cases, these devices are however placed at locations where they cannot be easily maintained, such as underground, underwater, or in remote locations where human operators are not available frequently. Therefore, it is crucial that such devices which usually run on batteries have a long lifetime so that a long service interval is possible. In this regard, extensive power measurements should be taken before a device is deployed in the field to get an overview about the power consumption so that the battery size can be adapted or the hard-/software of the device can be tuned in order to minimize the power consumption.

Since the supply voltage of the device is in general constant, the power measurement reduces to a current measurement. IoT devices are designed to go to sleep when not sensing or transmitting data, the sleep current is reduced to a few microamperes in order to maximize battery lifetime. Multimeters are not designed to measure such small currents with high precision or resolution. Since multimeters internally measure a voltage, currents are converted using a shunt resistor where a voltage drop a.k.a. burden voltage occurs which can be a few hundred millivolts, dependent on the most sensitive measurement range of the instrument. Using oscilloscopes for the measurements is attractive in order to easily get a graphical display of the current, however such devices only measure voltages, again in the region of a few hundred millivolts.

Because of the aforementioned restrictions of measurement instruments, it is desired to convert even very small currents into a voltage which is sufficiently high to be properly measured, while at the same time keeping the burden voltage in the region of a few millivolts, in contrast to hundreds of millivolts required by normal measurement devices as previously said. For this purpose, source measurement units (SMUs) are available which measure the voltage drop across a shunt and use a precision amplifier so that a small burden voltage is sufficient for the measurement and the resistance of the shunt can be kept small.

Since we widely work with IoT devices in our department, we need to frequently measure current intakes and therefore had a comprehensive survey on SMUs which we present in this report.

\section{Tested Devices}

There are two basic types of SMUs: Add-on devices provide an analog output voltage. Therefore, the time resolution of the measurement depends on the speed of the A/D converter which is built into the external instrument. For bench-top multimeters, the rate is a few hundred samples per second, whereas for oscilloscopes, the rate may be up to the gigasamples per second range.

Standalone SMUs include an A/D converter and a display in order to output the measured values. The time resolution of the measurement depends on the sample rate of the device's ADC.

In the following overview, some notable features of the tested SMUs are highlighted. The technical specifications are available in Table \ref{tab:Devices}.

\subsection{Add-on devices}

The \textbf{\textmu{}Current GOLD} available from the EEVblog online store\footnote{\href{https://eevblog.store/collections/test-equipment}{eevblog.store/collections/test-equipment}, Accessed: 2026-01-23} is an add-on device which converts the measured current into a voltage with a range large enough to be properly measured by an external meter. It features a pair of banana sockets for the current to be measured and another pair for the output voltage. For measurements with an oscilloscope, a banana-to-BNC adapter is required. A switch allows the selection between three different measurement ranges.

The device is powered by a 3\,V coin cell. There is an indicator when the battery voltage drops too low in order to ensure reliable measurements.

It should be noted that the device does not use fuses as overload protection for the current measurement in order to avoid an additional voltage drop. This however means that the shunt resistor used for the current measurement may be damaged if the measured current is too high.

The design of the \textmu{}Current GOLD is open-source\cite{Jones2009}.

A follow-up to the \textmu{}Current GOLD is the \textbf{tinyCurrent} available e.g. in the n-fuse store\footnote{\href{https://www.n-fuse.co/devices/tinyCurrent-precision-low-Current-Measurement-Shunt-and-Amplifier-Device.html}{www.n-fuse.co/devices/tinyCurrent-precision-low-Current-Measurement-Shunt-and-Amplifier-Device.html}, Accessed: 2026-01-23}. An important enhancement w.r.t. the \textmu{}Current is an additional BNC connector for the output voltage to connect an oscilloscope through a coaxial cable without additional adapter. The shielding of the coaxial cable is important when measuring signals with high time resolution because interference is reduced. There is also a version with an inverted BNC connector available which connects directly to the oscilloscope without a cable in between.

A second novelty of the tinyCurrent is a connector for an external power supply, which is useful for long-term measurements where the coin-cell battery would not suffice. 
The tinyCurrent is also Open Source\cite{tinyGit}.

\subsection{Stand-alone devices}

The \textbf{CurrentRanger} is again a further development of the tinyCurrent. It is extended by automatic measurement range selection, A/D conversion, data logging and an OLED display so that it also can be operated without an external measurement device. It is powered by a rechargeable lithium polymer (LiPo) battery.

The software is open source which is useful, e.g., if the automatic range selection should be tuned. This can happen if the device under test has a rapidly changing current intake. If the current gets too high because of a sudden increase of the current, the device runs into a brown-out until the shunt resistor is changed. Altering the software source code allows to change the speed at which the measurement range should be changed or to disable it altogether. Also, for long-term measurements, the automatic power-off function can be disabled.

A USB port allows for the connection of the device to a computer for data logging. One should use the USB port with care, since the ground of the USB connector is not separated from the rest of the circuitry. Wrong usage may damage the CurrentRanger, therefore inserting a USB isolator between the Current Ranger and the computer is recommended\cite{CRanger}.

The \textbf{\textmu{}SMU} is a four-quadrant device and therefore suitable both as a source and sink for positive and negative currents.
The output can be read through a USB interface; again a USB isolator may be needed dependent on the usage conditions as mentioned above.
The device is open source, Python code to control the unit from the computer is available on GitHub\footnote{\href{https://github.com/joeltroughton/uSmU}{github.com/joeltroughton/uSmU}, Accessed: 2026-01-23}.

\newpage

The \textbf{Nordic Semiconductors Power Profiler Kit 2} includes a power source which can be used to provide a supply voltage between 0.8 and 5\,V for the device under test (DUT). The device is controlled by the Software ``nRF Connect for Desktop" which is available for Linux, Windows and MacOS platforms. In source meter operation, an adjustable output voltage is provided for the DUT; in ampere meter operation, an external power supply can be used for the DUT. The maximum DUT current differs between the two modes -- 1000\,mA and 600\,mA, respectively. 

The \textbf{Otii Arc Pro} is another device with a builtin supply providing 0.5--5\,V voltage and 12\,W output power; as a special property, the power supply output for the DUT does not suffer from the burden voltage because the latter is actively compensated. It features two banana sockets, a USB port and an input for external power supplies. The device can also work as a current sink when used together with the optional Otii toolbox. The device is controlled by the Otii software which in the basic version is available free of charge for Linux, Windows and MacOS (Intel and Apple silicon). If scripting or a battery simulation by the device is desired, the respective software addons can be obtained by purchasing a permanent license or by a subscription.

A larger variant of the Otii Arc Pro is the \textbf{Otii Ace Pro} which features an isolated power supply, a higher output voltage of 25\,V, output power up to 30\,W and an increased sample rate on all channels. A complete list of enhancements is given in \cite{qoitech_comparison}.

The \textbf{Joulescope JS220} requires an external power supply for the DUT. For control, it is connected to a computer which has to run the software ``Joulescope UI'' which is readily packaged for Ubuntu Linux, Windows and Mac (Intel and Apple silicon). The software which is open source and written in Python can be downloaded from GitHub where also application examples are available. There are also third-party plugins available \cite{JoulescopeUI}.

The front panel of the device can be exchanged to provide different sets of inputs or outputs dependent on the application. A BNC connector allows for the synchronization with external measurement equipment. Moreover, GPIO ports enable the synchronization with the DUT. The device's physical sample rate of 2\,MSa/s is internally converted to 1\,MSa/s because of the limitations of the USB\,2 data output deployed in the device.

The \textbf{Rohde \& Schwarz NGU201} is ``fully'' standalone, i.e. unlike the other stand-alone devices, it does not need a computer for reading the data and external control of the device. Measured data can be stored on an external USB device. As with the JouleScope, digital inputs allow synchronization with the DUT. Due to the high price, the device is more targeted at labs and enterprise usage. Software add-ons support e.g. battery simulations.

\begin{landscape}
\begin{table}
	\centering
	\begin{adjustbox}{width=1\columnwidth}
		\small

		\begin{tabular}{@{}llllllllll@{}}
			
            \textbf{Device (Type)}                        & \textbf{Measurement Range}                                                                        & \textbf{Accuracy}                                                                              & \textbf{Burden Voltage}            & \textbf{Sample Rate}                            \\ \midrule
			Otii Ace Pro             & current range                                                                        & current range                                                                            & no burden voltage       & up to \SI{50}{\kilo\samples/\second} \\
			\cite{OttiManual}                    & \SI{50}{\nano\ampere} -- \SI{5}{\ampere}                                                     & \textpm(0.05\% + \SI{25}{\nano\ampere})                                                  &                            &                                       \\
			(stand-alone) & voltage range                                                                     & voltage range                                                                      &                            &                                       \\
			& \SI{0}{\volt} -- \SI{25}{\volt}                                                          & \textpm(0.01\% + \SI{1}{\milli\volt})                                                    &                            &                                       \\ \midrule
			Otii Arc Pro             & current range                                                                        & current range                                                                          & no burden voltage      & up to 4 kSa/s \\
			\cite{OttiManual}                    & \SI{50}{\nano\ampere}  -- \SI{5}{\ampere}                                                    & \textpm(0.1\% + \SI{50}{\nano\ampere})                                                   &                            &                                       \\
			(stand-alone) & voltage range                                                                     & voltage range                                                                       &                            &                                       \\
			& \SI{0.5}{\volt} -- \SI{5}{\volt}                                                         & \textpm(0.01\% + \SI{1}{\milli\volt})                                                    &                            &                                       \\ \midrule
			Nordic Power Profiler Kit 2          & \SI{200}{\nano\ampere} to \SI{1}{\ampere} (5 resistors)                              &                                                                                           & --                         & \SI{100}{\kilo\samples/\second}       \\
			\cite{NPPKGuide}         & (\SI{50}{\milli\ampere} – \SI{1}{\ampere})            R5 Res: \SI{1000}{\micro\ampere}  & R1 \textpm10\%   (\SI{100}{\nano\ampere} -- \SI{50}{\micro\ampere}) R Offset \textpm 2\% &                            &                                       \\
			(stand-alone) & (\SI{5}{\milli\ampere}  – \SI{50}{\milli\ampere})     R4 Res: \SI{50}{\micro\ampere}    & R2 \textpm10\%   (\SI{50}{\nano\ampere} -- \SI{500}{\micro\ampere}) R Offset \textpm 2\% &                            &                                       \\
			& (\SI{500}{\micro\ampere} – \SI{5}{\milli\ampere})     R3 Res: \SI{5}{\micro\ampere}     & R3 \textpm10\%   (\SI{500}{\micro\ampere} -- \SI{5}{\milli\ampere}) R Offset \textpm 2\% &                            &                                       \\
			& (\SI{50}{\micro\ampere} – \SI{500}{\micro\ampere})    R2 Res: \SI{0.5}{\micro\ampere}   & R4 \textpm10\%   (\SI{5}{\milli\ampere} -- \SI{50}{\milli\ampere})  R Offset \textpm 2\% &                            &                                       \\
			& (\SI{200}{\nano\ampere} – \SI{50}{\micro\ampere})    R1 Res: \SI{0.2}{\micro\ampere}   & R5 \textpm15\%   (\SI{50}{\milli\ampere} -- \SI{1}{\ampere}) R Offset \textpm 5\%        &                            &                                       \\ \midrule
			Joulescope JS220                     & current range                                                                          & current range                                                                          & --                         & \SI{2}{\mega\samples/\second}         \\
			\cite{JoulsJS220Guide}   & \SI{10}{\ampere} (max current \SI{3}{\ampere}) Res: \SI{175}{\micro\ampere}          & \SI{10}{\ampere}:    \textpm 0.25\%   \textpm \SI{1.5}{\milli\ampere}                     &                            &                                       \\
			(stand-alone) & \SI{180}{\milli\ampere} Res: \SI{15}{\micro\ampere}                                     & \SI{180}{\milli\ampere}:    \textpm 0.25\%   \textpm \SI{150}{\micro\ampere}              &                            &                                       \\
			& \SI{18}{\milli\ampere} Res: \SI{1.5}{\micro\ampere}                                     & \SI{18}{\milli\ampere}:    \textpm 0.25\%   \textpm \SI{15}{\micro\ampere}                &                            &                                       \\
			& \SI{1.8}{\milli\ampere} Res: \SI{150}{\nano\ampere}                                     & \SI{1.8}{\milli\ampere}:   \textpm 0.25\%   \textpm \SI{1.5}{\micro\ampere}               &                            &                                       \\
			& \SI{180}{\micro\ampere} Res: \SI{15}{\nano\ampere}                                      & \SI{180}{\micro\ampere}:    \textpm 0.25\%   \textpm \SI{150}{\nano\ampere}               &                            &                                       \\
			& \SI{18}{\micro\ampere} Res: \SI{1.5}{\nano\ampere}                                      & \SI{18}{\micro\ampere}:    \textpm 0.25\%   \textpm \SI{30}{\nano\ampere}                 &                            &                                       \\
			& voltage range                                                                    & voltage range                                                                      &                            &                                       \\
			& \SI{15}{\volt} Res: \SI{1.2}{\milli\volt}                                               & \SI{15}{\volt}:  \textpm 0.1\% \textpm \SI{10}{\milli\volt}                               &                            &                                       \\
			& \SI{2}{\volt} Res: \SI{180}{\micro\volt}                                                & \SI{2}{\volt}: \textpm 0.1\% \textpm \SI{2}{\milli\volt}                                  &                            &                                       \\ \midrule
			\textmu SMU                          & current range                                                                        & current resolution                                                                   & --                         & \SI{20}{\samples/\second}             \\
			\cite{usmu}              & --$50$ to +\SI{50}{\milli\ampere}                                                         & $\sim$\SI{10}{\nano\ampere}                                                               &                            &                                       \\
			(stand-alone) & voltage range                                                                    & voltage resolution                                                                       &                            &                                       \\
			& --$5$ to \SI{5}{\volt} (minimum step \textless{}1\,mV)                            & $\sim$\SI{0.6}{\milli\volt}                                                               &                            &                                       \\ \midrule
			Rohde \& Schwarz NGU201                       & current range                                                                     & current range (examples)                                                                  & --                         &  \SI{500}{\kilo\samples/\second}      \\
			\cite{RS_NGU_UserManual2}  & $\leq$ \SI{6}{\volt}:  \SI{8}{\ampere}                                                  & \SI{100}{\milli\ampere}: $\pm$0.025\% $\pm$\SI{15}{\micro\ampere}                                        &                            &                                       \\
			(stand-alone) & \textgreater  \SI{6}{\volt}:  \SI{3}{\ampere}                                           &       \SI{3}{\ampere}: $\pm$0.025\% $\pm$\SI{250}{\micro\ampere}                                                                                    & --                         &                                       \\
			& voltage range                                                                    & voltage range (example)                                                                      &                            &                                       \\
			& 0 to \SI{6}{\volt}, 0 to \SI{20}{\volt}                                                                   & \SI{6}{\volt}: $\pm$0.02\% $\pm$\SI{500}{\micro\volt}                                        &                            &                                       \\ \bottomrule
			
			µCurrent Gold\cite{Jones2009}              & three current ranges                                                               & \textless \textpm 0.1\% in \si{\milli\ampere} range                          & \SI{20}{\micro\volt} / \si{\milli\ampere} in \si{\milli\ampere} range    & dependent on oscilloscope     \\
			/tinyCurrent  \cite{tinyGit}                                    & \textpm 1.250 / \SI{2.550}{\milli\ampere}                                            & \textless \textpm 0.05\% in \si{\micro\ampere}- and \si{\nano\ampere} range  & \SI{10}{\micro\volt}\ / \si{\micro\ampere} in \si{\micro\ampere} range   & / device                  \\
			(add-on) & \textpm 1.250 / \SI{2.550}{\micro\ampere}                                            &                                                                                 & \SI{10}{\micro\volt} / \si{\nano\ampere}   in \si{\nano\ampere}  device   &                              \\
			& \textpm 1.250 / \SI{2.550}{\nano\ampere}                                             &                                                                                 &                                                                            &                              \\ \midrule
			Current Ranger \cite{CRanger}                                      & three measurement ranges in two modes                                                       & \textpm 0.1\% (\si{\milli\ampere})                                            & \SI{17}{\micro\volt} / \si{\milli\ampere} im \si{\milli\ampere} range    & dependent on  oscilloscope     \\
			(stand-alone/add-on) & Bidirectional mode                                                                   & \textpm 0.05\% (\si{\micro\ampere}, \si{\nano\ampere})                        & \SI{10}{\micro\volt}\ / \si{\micro\ampere} im \si{\micro\ampere} range   & / measurement device                  \\
			& \textpm 0 to 1.650 \si{\milli\ampere} / \si{\micro\ampere} / \si{\nano\ampere}          &                                                                                 & \SI{10}{\micro\volt} / \si{\nano\ampere}  in \si{\nano\ampere}  range    &                              \\
			& Unidirectional mode                                                                  &                                                                                 &                                                                            &                              \\
			& 0 to 3.300 \si{\milli\ampere} / \si{\micro\ampere} / \si{\nano\ampere}                  &                                                                                 &                                                                            &                              \\ \bottomrule
		\end{tabular}
	\end{adjustbox}
	\vspace{6pt}
	\caption{Specifications of measurement devices discussed in this report
		}
	\label{tab:Devices}
\end{table}
\end{landscape}

\section{Example Measurements}

In our department, we develop a sensor board named MoleNet\footnote{\href{https://www.molenet.org}{www.molenet.org}, Accessed: 2026-01-23} which features an ESP32-S3 microcontroller, along with various interfaces to connect different types of sensors: UART (Universal Asynchronous Receiver and Transmitter) has been used in the computer communications area since an extended period of time. SDI-12 (Serial Digital Interface) is similar to UART concerning the signaling protocol, but has some different electrical characteristics. It is widely used by environmental sensors. Finally, the I$^\textrm{2}$C (Inter-Integrated Circuit) interface is provided which was originally designed to connect ICs on a printed circuit board, but is also used by some sensors. 

The MoleNet board also provides the supply voltage for the connected sensors. Furthermore, the board includes WLAN and LoRa interfaces for communication and an SD card reader for data logging. The SD card uses the SPI (Serial Peripheral Interface) for the communication with the microcontroller.

The board can be programmed using Arduino or MicroPython. We used the latter for the energy consumption tests described in this report. In the course of the development, it became necessary to measure the power consumption of the board due to the reasons of constrained battery supplies in typical sensor nodes as discussed in the introduction section. Our selection of the source measurement unit therefore was guided by the current intake of the MoleNet board, flexibility of using the software and affordable cost. 

The upper limit of the MoleNet board's current intake is determined by the voltage regulator on the board, which is a Texas Instruments TLV731 with a max. current of 1\,A. The reason to select a voltage regulator with a relatively high current is that the board also should be able to power the sensors connected to it. This requirement ruled out using the \textmu{}SMU for the measurements which is specified with a maximum load of 50\,mA only. Furthermore, we preferred using an SMU which comes with open-source software and thus allows to alter the software according to varying needs and saves cost as well. Therefore, we decided to select the tinyCurrent as the add-on device which suited our needs best and the JouleScope as the standalone device. For reference, we also tested the Rohde \& Schwarz NGU201.

\subsection{Test suite}
We ran a number of tests with different peripherals connected to the board as well as when the board is put into sleep mode. The tests were executed in an automated way by a program running on the ESP32 microcontoller of the MoleNet board. In all tests described in this section, the current intake was measured between the voltage regulator and the MoleNet board's circuitry. The individual testing steps included activity of varying peripheral devices and sleep phases as well, as follows:

\begin{itemize}
	\item LoRa: Send the string ``Hello World'' through the LoRa interface using a transmit power of +17\,dBm, a spreading factor of 12 and a bandwidth of 125\,kHz.
	\item SDI-12 and UART: A 5TM volumetric water content (VWC) sensor used for soil monitoring was connected to the MoleNet board using the SDI-12 interface. The sensor was initialised, after that, a reading was taken whose value was written to the serial console. No further data processing took place.
	\item microSD card: the frequency of the SPI interface used to control the card was set to 100\,kHz. The card was initialized, the files contained on the card were read out and another text file was written and read. After that, the card was deinitialized and ejected. Different SD card models may have different current intakes, so for comparability, for all experiments in this report, a SanDisk Class 10 card with 8\,GB capacity was used.
	\item A BME280 air sensor was connected though the I$^\textrm{2}$C interface which runs at a frequency of 10\,kHz. Values for air pressure, temperature and humidity were read out.
	\item An idle phase where the microcontroller remained powered up but does not execute any command, as well as the Light Sleep and Deep Sleep modes were tested for a period of 30 seconds, respectively.
\end{itemize}

After each individual test, the ESP32 microcontroller waits for pressing a pushbutton before the next test is executed. Furthermore, the ESP32 provides a trigger signal on a dedicated pin on the MoleNet board to indicate the beginning and end of individual tests. This functionality is used by the SMUs under test to halt recording while waiting for the start of the next test.

We ran the complete test sequence with the JouleScope and the R\&S meter, but not with the tinyCurrent, because the latter does not support a trigger input signal for control purposes and was hence not used at this point.

\begin{figure*}
	\includegraphics[width=0.98\textwidth]{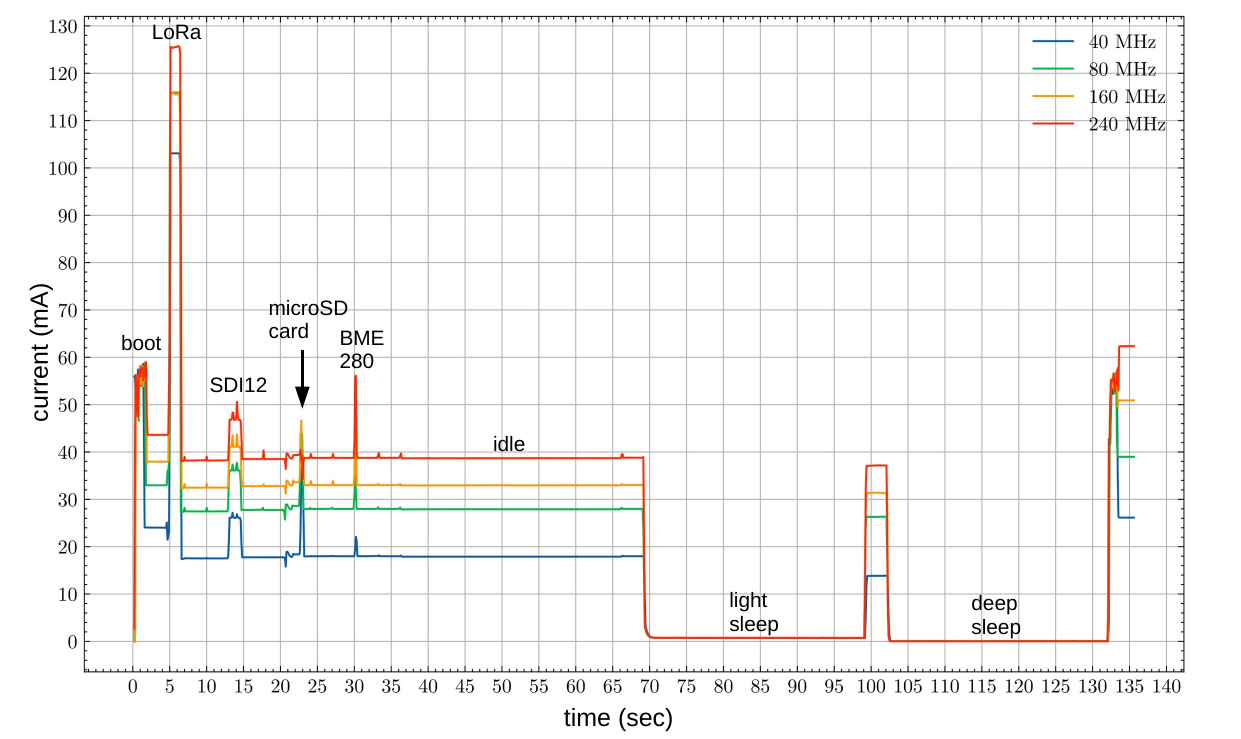}
	\caption{Current intake of MoleNet board vs.\ time, measured with R\&S for different microcontroller clock frequencies}
	\label{fig:rs}
\end{figure*}

\begin{figure*}
    \includegraphics[width=\textwidth]{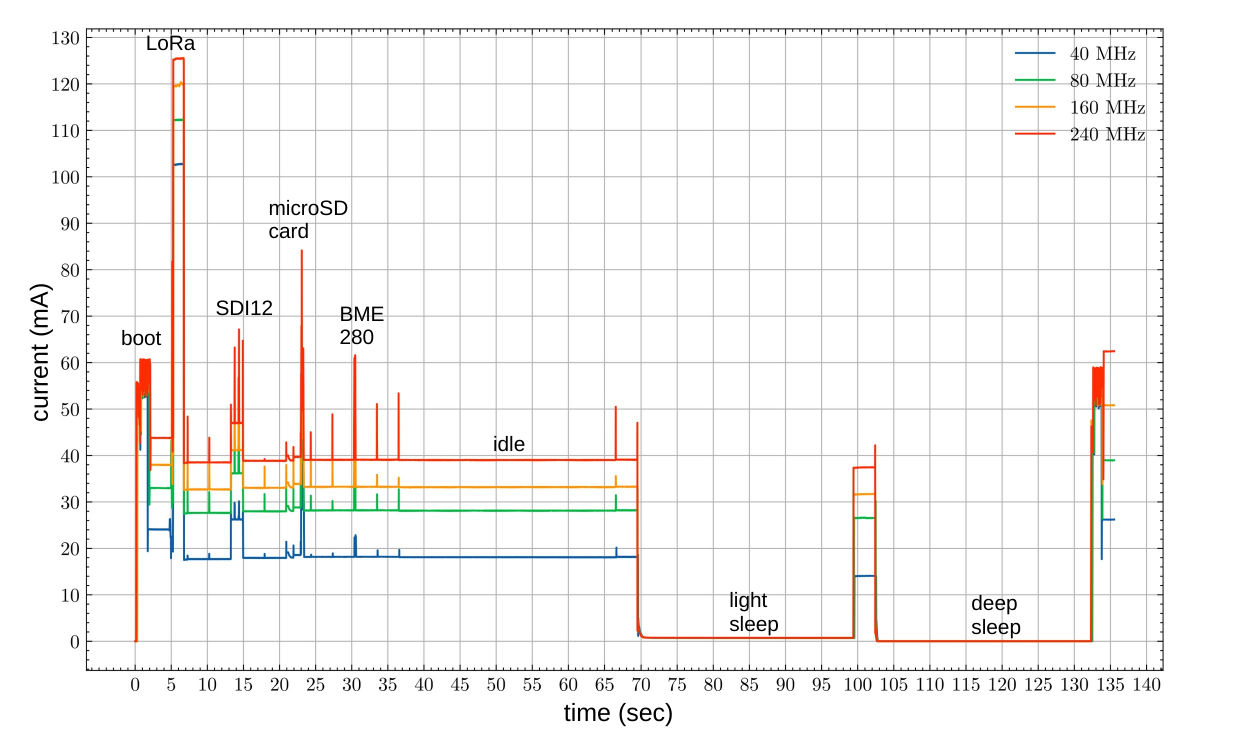}
    \caption{Current intake of MoleNet board vs.\ time, measured with JouleScope for different microcontroller clock frequencies}
    \label{fig:joulescope}
\end{figure*}

The measurement results obtained are shown in Figs.~\ref{fig:rs} and \ref{fig:joulescope}. While in the time periods with constant current the readings are identical as expected, there are differences when short current spikes appear. The JouleScope shows all these short pulses whereas with the R\&S, they are flattened. This observation suggests that the R\&S's sample rate is too slow. We further investigate this issue in the following section.

\subsection{Individual test with LoRa}

Besides the test suite described above, we ran an individual current measurement where only the LoRa interface is active. On the one hand, this allowed to include the tinyCurrent SMU into the test -- as described before, it could not be deployed in the test suite due to lack of a trigger signal input. In the LoRa-only test, no trigger signal is needed, since only a single setup is tested. On the other hand, taking a third device into account is also useful to clarify the curve smoothing with the R\&S device as discussed before. In order to achieve the latter goal, we also increased the sampling rate w.r.t. the test suite to get a more detailed output. This worked without problems with the tinyCurrent where the signal is anyway sampled by the external device, and with the JouleScope. However, issues occured with the R\&S which provides a function called Fast Logging with an increased sample rate. However, trying to activate the function regularly resulted in a system message reporting that Fast Logging was disabled because of an internal critical error, although the latest firmware version was installed. Contacting the support helpdesk also did not yield helpful information. Further, trying to enable Fast Logging through an external control software provided by the vendor resulted in the same error. Hence, the R\&S could only be used with the default sample rate of 10\,Hz.

\begin{figure}
	\begin{center}
	\includegraphics[width=0.428\textwidth]{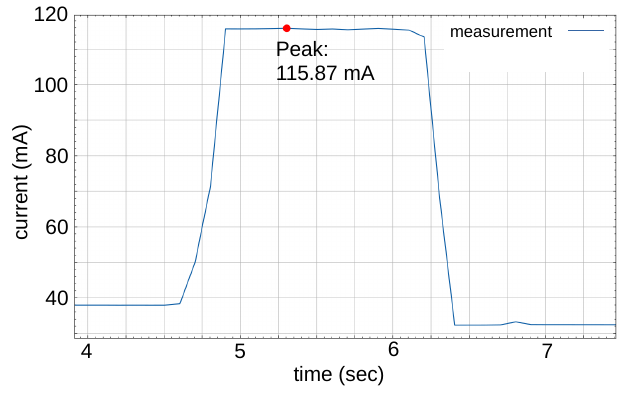}
	\caption{Current intake of MoleNet board vs.\ time for running the LoRa interface, measured with R\&S}
	\label{fig:lora_rs}
	\end{center}
\end{figure}

\begin{figure}
	\begin{center}
	\includegraphics[width=0.428\textwidth]{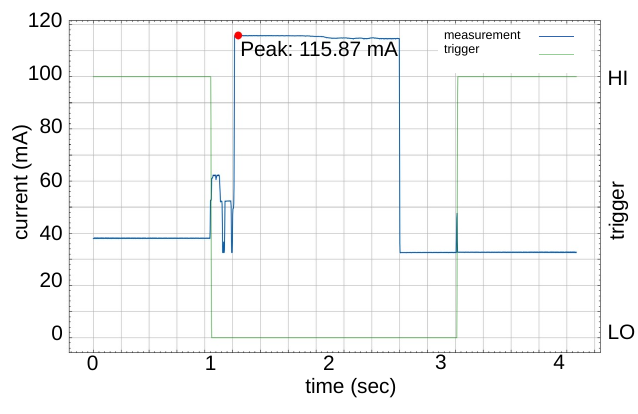}
	\caption{Current intake of MoleNet board vs.\ time for running the LoRa interface, measured with JouleScope}
	\label{fig:lora_joulescope}
    \end{center}
\end{figure}

\begin{figure}
	\begin{center}
	\includegraphics[width=0.423\textwidth]{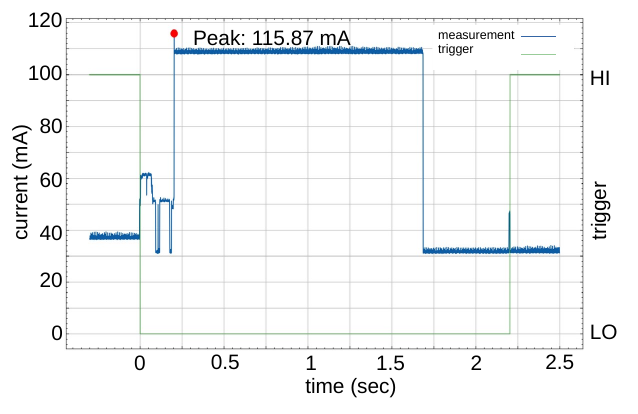}
	\caption{Current intake of MoleNet board vs.\ time for running the LoRa interface, measured with tinyCurrent}
	\label{fig:lora_tinycurrent}
    \end{center}
\end{figure}

The results are shown in Fig.~\ref{fig:lora_rs}, \ref{fig:lora_joulescope} and \ref{fig:lora_tinycurrent} for R\&S, JouleScope and tinyCurrent, respectively. In the R\&S measurement, only a single peak is visible with smooth edges, whereas in the JouleScope and tinyCurrent measurements the peak has sharp edges and further pulses and spikes are visible, which confirms that the measurements are taken at a higher sample rate by the JouleScope and tinyCurrent, whereas with the R\&S, details of the signal are lost due to the limited sample rate.

\section{Conclusion and Outlook}

In this work, we tested different measurement devices using the MoleNet board as a demonstration example.
The tinyCurrent provides a low-cost and easy-to-use extension to run low-current measurements with a multimeter or oscilloscope. However, due to the lack of automatic range selection, it cannot be used if the measured current varies between high and very small values; also, due to the lack of a trigger input, the suitability for automated measurement suites is limited. The R\&S meter could not be run with the fast logging mode. Therefore, the JouleScope appears to be the most suitable candidate among the SMUs tested in this report.

Further tests should investigate on the energy consumption of the LoRa module which is highly dependent on the selected transmit power. Moreover, when the microcontroller has to perform computationally expensive tasks, it should be investigated if increasing the clock can reduce the energy consumption, since on the one hand the power intake is higher, but on the other hand the computations require less time.

\bibliography{bibliography}
\bibliographystyle{plain}

\end{document}